\documentclass[twocolumn]{aastex62}

\usepackage{apjfonts}
\usepackage[T1]{fontenc}
\usepackage{color}
\usepackage{amsmath,amstext}
\usepackage{hyperref}
\usepackage{natbib}
\usepackage{graphicx}
\usepackage{gensymb}

\begin{document}

\title{\sc Propagation of a realistic magnetar jet through binary neutron star merger medium and implications for short gamma-ray bursts}
\author[0000-0002-6569-5769]{Gustavo Soares}
\affil{Department of Physics, Oregon State University, 301 Weniger Hall, Corvallis, OR 97331, USA}
\correspondingauthor{Gustavo Soares}
\email{rodrgust@oregonstate.edu}
  
\author{Pablo Bosch}
\affil{GRAPPA, Anton Pannekoek Institute for Astronomy and Institute of High-Energy Physics, University of Amsterdam, Science Park 904, 1098 XH Amsterdam, The Netherlands}

\author[0000-0002-9190-662X]{Davide Lazzati}
\affil{Department of Physics, Oregon State University, 301 Weniger Hall, Corvallis, OR 97331, USA}
  
\author[0000-0002-9371-1447]{Philipp M\"osta}
\affil{GRAPPA, Anton Pannekoek Institute for Astronomy and Institute of High-Energy Physics, University of Amsterdam, Science Park 904, 1098 XH Amsterdam, The Netherlands}

\begin{abstract}
The origin of short gamma-ray bursts (sGRBs) is associated with outflows powered by the remnant of a binary neutron star merger. This remnant can be either a black hole or a highly magnetized, fastly spinning neutron star, also known as a magnetar. Here, we present the results of two relativistic magnetohydrodynamical (RMHD) simulations aimed at investigating the large-scale dynamics and propagation of magnetar collimated outflows through the medium surrounding the remnant. The first simulation evolves a realistic jet by injecting external simulation data, while the second evolves an analytical model jet with similar properties for comparison. We find that both outflows remain collimated and successfully emerge through the static medium surrounding the remnant. However, they fail to attain relativistic velocities and only reach a mean maximum speed of $\sim 0.7c$ for the realistic jet, and $\sim 0.6c$ for the analytical jet. We also find that the realistic jet has a much more complex structure. The lack of highly relativistic speeds, that makes these jets unsuitable as short GRB sources, is due to numerical limitations and not general to all possible magnetar outflows. A jet like the one we study, however, could give rise to or augment a blue kilonova component. In addition, it would make the propagation of a relativistic jet easier, should one be launched after the neutron star collapses into a black hole.

\end{abstract}
\section{Introduction}
\label{sec:intro}

The merger of binary neutron stars (BNS) leads to the formation of another compact object. Its final nature will depend on factors such as the remnant's mass and its ability -- or lack thereof -- to support itself against its own gravity as it spins down and cools off (see \citealt{Radice2020review} for a recent review of BNS mergers). The most massive remnants will immediately collapse into a black hole, while the less massive ones will be neutron stars. This latter case may be further subdivided into unstable hypermassive (HMNS) or supramassive neutron stars \citep{Baumgarte2000,Cook1992,Cook1994} -- both of which will invariably undergo gravitational collapse into a black hole -- and a stable neutron star.

These events are also among the most luminous in the universe \citep{Abbott2017b}. They have been long hypothesized to be the origin of short gamma-ray bursts (sGRBs; \citealt{Eichler1989,Narayan1992}), and play a fundamental role in the origin of kilonovae (KN) and the formation of heavy elements -- see e.g. \cite{Metzger2019review} for a review.

The prompt gamma-ray emission of GRBs is primarily attributed to two mechanisms. The first of these is internal shocks between shells within relativistic jets powered by the central engine \citep{Narayan1992,Paczynski1994,Rees1994}, while the second is external shocks between the leading shells and the surrounding interstellar medium \citep{Meszaros1992,Rees1992,Katz1994}. Additionally, both mechanisms might be necessary to explain some observations \citep{Piran1998}. Nevertheless, in order to power such jets, a compact object such as a black hole or a neutron star is required, along with strong magnetic fields.

The multimessenger observation of GW170817 and its electromagnetic counterparts \citep{Abbott2017b,Abbott2017a} confirmed many of the predictions highlighted above, and placed significant constraints on other hypotheses. Yet, there still remain a few unanswered questions, in particular regarding the origin and engine powering the associated gamma-ray burst GRB 170817A, which was first detected 1.7 seconds after the initial gravitational wave detection \citep{Abbott2017b}.

Follow-up observations of the KN associated to GW170817 highlighted tensions with simulation results with respect to the amount of KN ejecta and its velocity. Moreover, in order to explain the observations, existing models often require additional constraints related to the NS remnant's lifetime, radius, and accretion disk mass (e.g. \citealt{Bauswein2013, Hotokezaka2013a,Fahlman2018}). Based on this, \cite{Metzger2018} proposed that this tension can be alleviated if the engine powering GRB 170817A is a rapidly spinning, strongly magnetized HMNS remnant with a lifetime of $t \sim 0.1-1\mathrm{s}$.

Multiple groups have explored the feasibility of a magnetar engine for sGRBs, but again a successful case is highly dependent on factors such as the remnant's lifetime and neutrino effects. For instance, \cite{Ruiz2016} found that a jet was produced only after the HMNS had further collapsed into a black hole, while stable magnetars were hampered by baryon pollution at polar regions \citep{Ciolfi2017,Ciolfi2019,Ciolfi2020}, which in turn prevented the formation of jets and could not be taken care of in the absence of neutrino effects in those simulations.

This approach, where a jet is launched during the HMNS phase, differs from other works where the jet is launched after the neutron star collapses into a black hole (\citealt{Nathanail2020,Nathanail2021,Gottlieb2022a}). Other works (e.g. \citealt{Rezzolla2011,Kiuchi2014}) capture the entire system evolution from the BNS merger, and the resulting HMNS quickly collapses into a BH, while \cite{MurguiaBerthier2017, MurguiaBerthier2021} discuss the features of GRBs triggered by BNS mergers and the jet properties, particularly in the case of GW170817.

\cite{Moesta2020}, hereafter M20, performed a series of high-resolution simulations of NS merger remnants with a nuclear equation of state, neutrino cooling and heating, while adding strong magnetic fields similar to those in magnetars. Due to the high-resolution being able to resolve MRI, M20 found that the strong magnetic fields in the HMNS were capable of launching jets when neutrino effects were included, in particular neutrino cooling which reduced baryon pollution around the poles and allowed the jets to reach Lorentz factors of $2-5$, depending on the simulation details.

Here, we evolve the outflowing material from the B15-low simulation from M20, which originated during the HMNS remnant stage, for another order of magnitude in distance, and compare those features to those of an analytical jet simulation with similar initial conditions. We find that the outflowing material reaches mean velocities of $\lesssim 0.7c$ in the realistic jet, and about $0.6c$ in the analytical jet, although we must consider numerical limitations affecting this result. The overall structure formed by the ejecta is that of a cocoon, consisting of a shocked jet (which might be relativistic) in the center surrounded by shocked stellar material. Within this cocoon, we found velocities of $\lesssim 0.5c$ in both simulations. Overall, although both jets remained collimated and achieved moderate to high velocities, we found that they are most likely unable to power sGRBs. However, this does not necessarily imply that magnetars are, as a whole, unsuitable sources of sGRBs, because full neutrino transport has not been included and the ejected material may be less baryon-rich if neutrino pair-annihilation is taken into consideration \citep{Fujibayashi2017}.

In this paper, we describe our simulations in Section \ref{sec:methods}. The main results are presented in Section \ref{sec:results}, and they are further discussed in Section \ref{sec:discussion}. We present our conclusions in Section \ref{sec:conclusions}.

\section{Simulations: setup and overview}\label{sec:methods}

We ran two 3D RMHD simulations with PLUTO 4.4 \citep{Mignone2007}. Our integration setup consists of a second order Runge-Kutta time-stepping, an HLLC Riemann solver, and piecewise parabolic reconstruction. We assumed an ideal gas equation of state for the background gas and enforced $\nabla\cdot\mathbf{B}=0$ through divergence cleaning.

The first of our simulations, which we will constantly refer as having a ``realistic jet'', is evolved with periodic injections of outflowing material from a general relativistic MHD simulation of a newly formed magnetar (M20, see Section \ref{sec:inputdata} for more details) mapped into our grid, as explained in Section \ref{sec:rd}. The second simulation is described in Section \ref{sec:analytical} and consists of an analytical jet injected at and around the grid inner boundary.

In both cases, we adopt a set of background initial conditions consisting of background gas with ambient density $\rho_a = 10^{-4}\;\mathrm{g\,cm^{-3}}$ and ambient pressure $p_a = 6\times10^{-7}c^2\mathrm{g\,cm^{-3}}$. We also added a semi-spherical material at rest with radially decreasing density and pressure according to a Gaussian profile,
\begin{equation}
    \rho(r) = \rho_i\exp{(-r^2/2r_*)},
    \label{eq:ejecta}
\end{equation}
where $r_* = 3.8\times10^8\;\mathrm{cm}$, $\rho_i$ is the density of a mass of $2\times10^{30}\;\mathrm{g}$ within a sphere of radius $r_*$, and $p(r) = 10^{-3}\rho(r)$.

For both our simulations, we adopted a Cartesian grid extending for $8\times10^8\;\mathrm{cm}$ in the $x-$ and $y-$directions, and $1.4\times10^9\;\mathrm{cm}$ in the $z$-direction.

\subsection{Input data}\label{sec:inputdata}

The data mapped into our realistic jet simulation is described in detail in M20. It originated from an ideal GRMHD simulation with an adaptive mesh with the open-source \texttt{Einstein Toolkit} \citep{Babiuc-Hamilton2019,Loeffler2012,Schnetter2004} module \texttt{GRHydro} \citep{Moesta2014}. Initial data in that simulation was mapped from a GRHD binary neutron star (BNS) merger simulation with \texttt{WhiskyTHC} \citep{Radice2012} $17\;\mathrm{ms}$ after the BNS merger. The details of this mapping can be seen in M20, but in summary, the mapping occured at $t_\mathrm{map} - t_\mathrm{merger} = 17\,\mathrm{ms}$, with a magnetic field given by $A_\mathrm{r} = A_\theta = 0; A_\phi = B_0r_0^3(r^3 + r_0^3)^{-1}r\sin\theta$, where $r_0 = 20\,\mathrm{km}$ and $B_0 = 10^{15}\,\mathrm{G}$, and an outer boundary of $\sim 355\,\mathrm{km}$.

The simulation in M20 employs the equation of state of \cite{Lattimer1991}, with $K_0 = 220\;\mathrm{MeV}$, along with the neutrino leakage and heating approximations of \cite{OConnor2010,Ott2012}. It tracks electron neutrinos, electron antineutrinos and heavy-lepton neutrinos, the latter being treated as a single species. Neutrino cooling is implemented by approximating energy-averaged neutrino optical depths followed by local estimates of energy and lepton loss rates, while neutrino heating is approximated using a prescription for the neutrino heating rate. It depends on the neutrino luminosity as predicted by the neutrino leakage approximation along radial rays, as well as the electron mass, the neutron mass, the speed of light, the rest mass density, the neutron (proton) mass fraction for electron neutrinos (antineutrinos), the mean-squared energy of the neutrinos, and the mean inverse flux factor. The explicit prescription and details can be found in M20. Finally, the mapped GRHD BNS remnant is endowed with an \emph{ad hoc} parameterized magnetic field prescribed through the vector potential with $A_r=A_{\theta}=0$ and $A_{\phi} = B_0r_0^{\phantom{0}3} (r_0^{\phantom{0} 3} + r^3)^{-1} r\sin\theta$. The parameters $B_0$ and $r_0$ control the overall strength and the falloff of the magnetic field, respectively. In M20 the numerical evolution was carried out, at three different resolutions, with values $B_0=10^{15} \mathrm{G}$, to have magnetar-level magnetic field strengths, and $r_0 = 20 \mathrm{km}$, to keep the magnetic field nearly constant within the HMNS. Additionally, to prevent violations of the divergence-free constraint of the magnetic field, $\nabla \cdot \mathbf{B} = 0$, a constrained transport scheme is employed. In this work, we map the first level of lowest resolution simulation B15-low from M20 onto our grid as described below, in Sec.~\ref{sec:rd}.

\subsection{Data injection}\label{sec:rd}

Our Cartesian grid for this simulation was constructed as follows. We kept a $237\times 237 \times 237$ box at the origin of our simulation, in which the jet was injected. This corresponds to a cube of $\sim 420\;\mathrm{km}$ in each direction. This was chosen to match the $237$ cells in the $xy$-plane in the original simulation (within their first level of AMR), and preserve the original jet as best as possible without too many interpolations into grids with drastically different sizes, which could lead to inaccuracies in the data as it was injected. Within this box, all the cells have the same size as in the original simulation, within one level of AMR,  with a resolution of $h=1.77\mathrm{km}$. We then proceeded to extend this original box along the $x$- and $y$-directions logarithmically, with $100$ cells on either side. Finally, we extended the $z$-direction by adding $600$ cells in a logarithmic grid. In total our grid has $437$ cells in the $x$- and $y$- directions, and $837$ cells in the $z$-direction.

The snapshots are extracted from M20 at 0.15 ms intervals during the (quasi) steady-state operation of the jet which lasts about 5 ms. For reference, in M20, the HMNS is evolved roughly 23 ms before collapsing to a black hole. In our current simulation, we inject these snapshots at the same rate as they were extracted, 0.15ms, keeping the time interval fixed. Once all the snapshots are injected we restart the procedure until the end of the simulation; effectively cycling over the available snapshots and treating the injection process as a loop. Given that we are looping over the (quasi) steady-state operation of M20 simulations, this scheme provides a reasonable approximation for a long-lived jet. The total evolution time is $\sim 67\mathrm{ms}$ (i.e. around 13 completed injection cycles), thus any specific features arising  from jet launching or the eventual collapse of the HMNS will not be reproduced here. Carrying out this simulation on a big domain for a relatively long time without reducing the dimensionality was made possible by the simplification of the equation of state and by being away from the compact object, allowing us not to be concerned about resolving it. This allowed us to use a more efficient special relativistic code and to significantly increase the time step. A similar technique was used by \cite{Pavan2021} and \cite{Lazzati2021}. In those works, however, an analytic jet was injected into a realistic ambient medium, while here we attempt the opposite feat by injecting a realistic jet into an analytic environment.

\begin{figure}
    \centering
    \includegraphics[width=\linewidth]{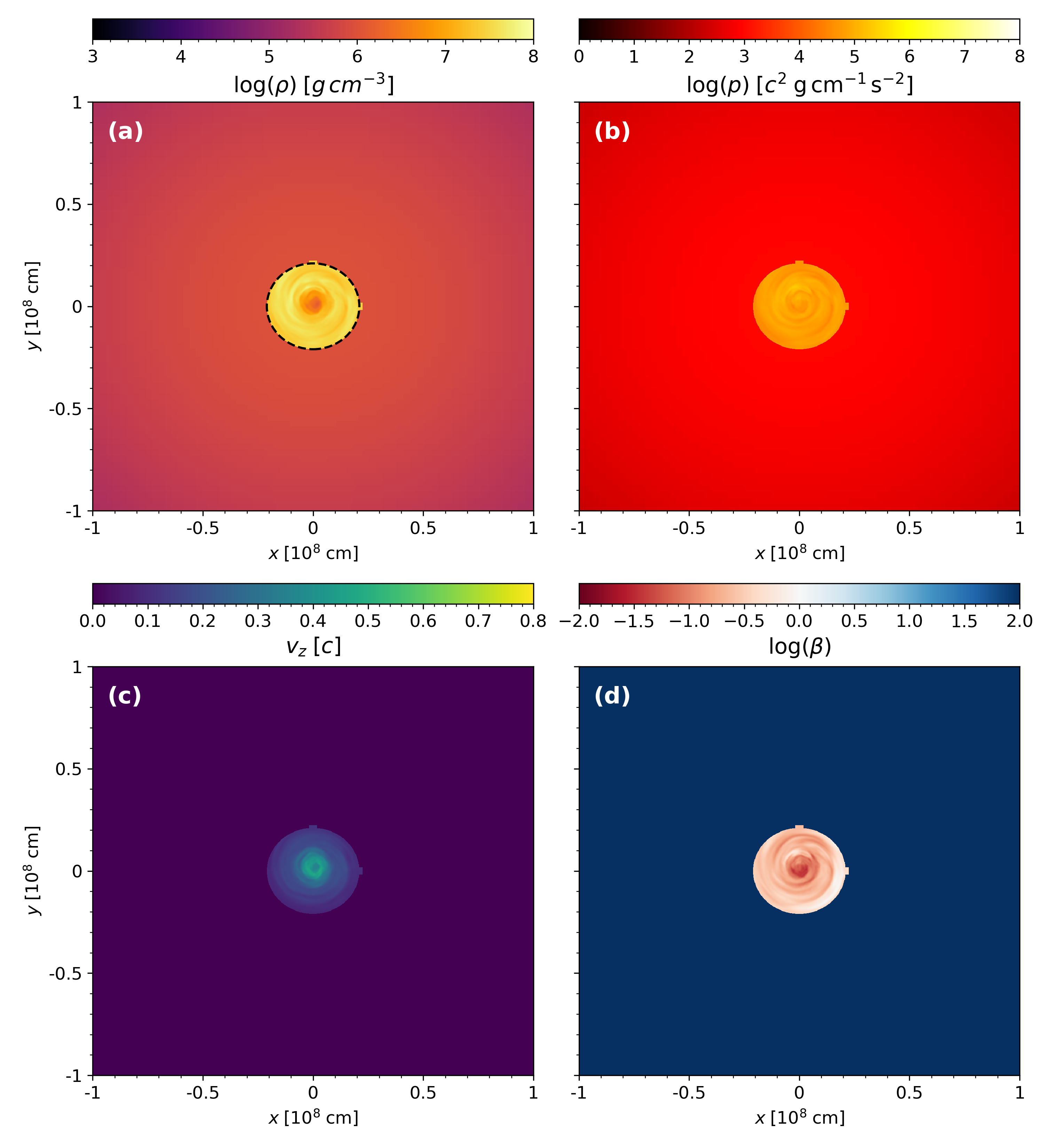}
    \caption{$xy$-plane cut at $t=0\,\mathrm{s}$ of the following quantities: (a) gas density $\rho$, (b) pressure $p$, (c) velocity $v_z$ along the $z$-direction, and (d) plasma $\beta$, all at $z=0$ at the beginning of the realistic jet simulation, zoomed in to focus on the material mapped from M20. Injection occurred within a radius of $0.21\times 10^8\;\mathrm{cm}$ from the $xy$-plane origin; this region is shown on panel (a) as a black dashed circle.}
    \label{fig:rj_xy0}
\end{figure}

Figure \ref{fig:rj_xy0} shows a few physical quantities of interest -- both background and injected -- in the equatorial plane for $z=0$, at the start of the simulation ($t = 0\,\mathrm{s}$). The four panels show (a) the gas density $\rho$, (b) the gas pressure $p$, (c) the velocity $v_z$ along the $z$-direction, and (d) the plasma $\beta$. The central material corresponds to the data mapped from M20 into our grid. Data mapping occurred within a circle of radius $0.21\times 10^8\;\mathrm{cm}$ -- this length was chosen because it corresponds to the first level of grid refinement in M20. Decreasing radially in the $xy$-plane are the extra density and pressure representing stellar material. Along the positive $x$- and $y$-directions as we transition from the injection region towards the rest of the simulation domain, there are two small bulging features on all panels. These are numerical in nature and likely occurred during data interpolation. They do not seem to affect the evolution of our simulation.

Upon looking at the panels in Figure \ref{fig:rj_xy0}, we can also note a hollow geometry in the jet. Its central axis is characterized by lower densities compared to the rest of the material as we move away from this axis (evidently discarding the low-density region outside of the injection area). This lower density along the jet's axis is accompanied by higher pressure -- which remains relatively constant within the injected area -- and low $\beta$, an indication of high magnetization.

\subsection{Analytical jet}\label{sec:analytical}

The background initial conditions used in this simulation are the same that were used in the realistic jet simulation. Our grid for this simulation differs slightly from that; here we have a total of $440$ cells in the $x-$ and $y-$directions, with $200$ of these concentrated within a radius $r_0$ in the $xy$-plane, taken to be the initial radius of the domain in which we injected our data in the previous simulation. Finally, there are $600$ cells in the $z-$direction. Our prescription for the jet follows closely that of \cite{Mignone2009,Mignone2013,Gottlieb2020a}, with a few modifications aimed at matching as much as possible the values of quantities in this simulation to those of our data-injected simulation. Within $r_0$, we assume a density $\rho_0$, which is taken to be the mean density in the injection domain of M20, and is then radially smoothed out as we multiply the density along the $xy$-plane by the profile
\begin{equation}
    \frac{1}{\cosh{(r/a)^8}},
    \label{eq:radial_profile}
\end{equation}
where $a = r_0/2$ and $r = \sqrt{x^2 + y^2}$. Within the jet, the purely toroidal magnetic field is given by $B_\phi(r) = \gamma b_\phi(r)$, where
\begin{equation}
    b_\phi = 
    \begin{cases}
    b_m r/a,& \text{if } r\leq a\\
    b_m a/r,& \text{if } a < r < r_0,
    \end{cases}
    \label{eq:bmag}
\end{equation}
and $b_m = \sqrt{-4p_0\sigma_\phi/[a^2(2\sigma_\phi - 1 + 4\log a)]}$. Here, $\sigma_\phi = 1$ is the toroidal magnetization parameter and $p_0$ was taken from the data-injected simulation, being the mean pressure within the injected domain, $p_0 = 7\times 10^{-5}\;c^2\,\mathrm{g\,cm^{-3}}$. Finally, the pressure inside the jet, $p_j$, is given by
\begin{equation}
    p_j(r) = p_0 + b_m^2[1 - \mathrm{min}(r^2/a^2, 1)],
\end{equation}
which is also multiplied by Eq. \ref{eq:radial_profile}. The jet's initial velocity is  $v_j \sim 0.4c$.

\section{Results}
\label{sec:results}

\subsection{Realistic jet}
\label{sec:results_rj}

\begin{figure*}
    \centering
    \includegraphics[width=\linewidth]{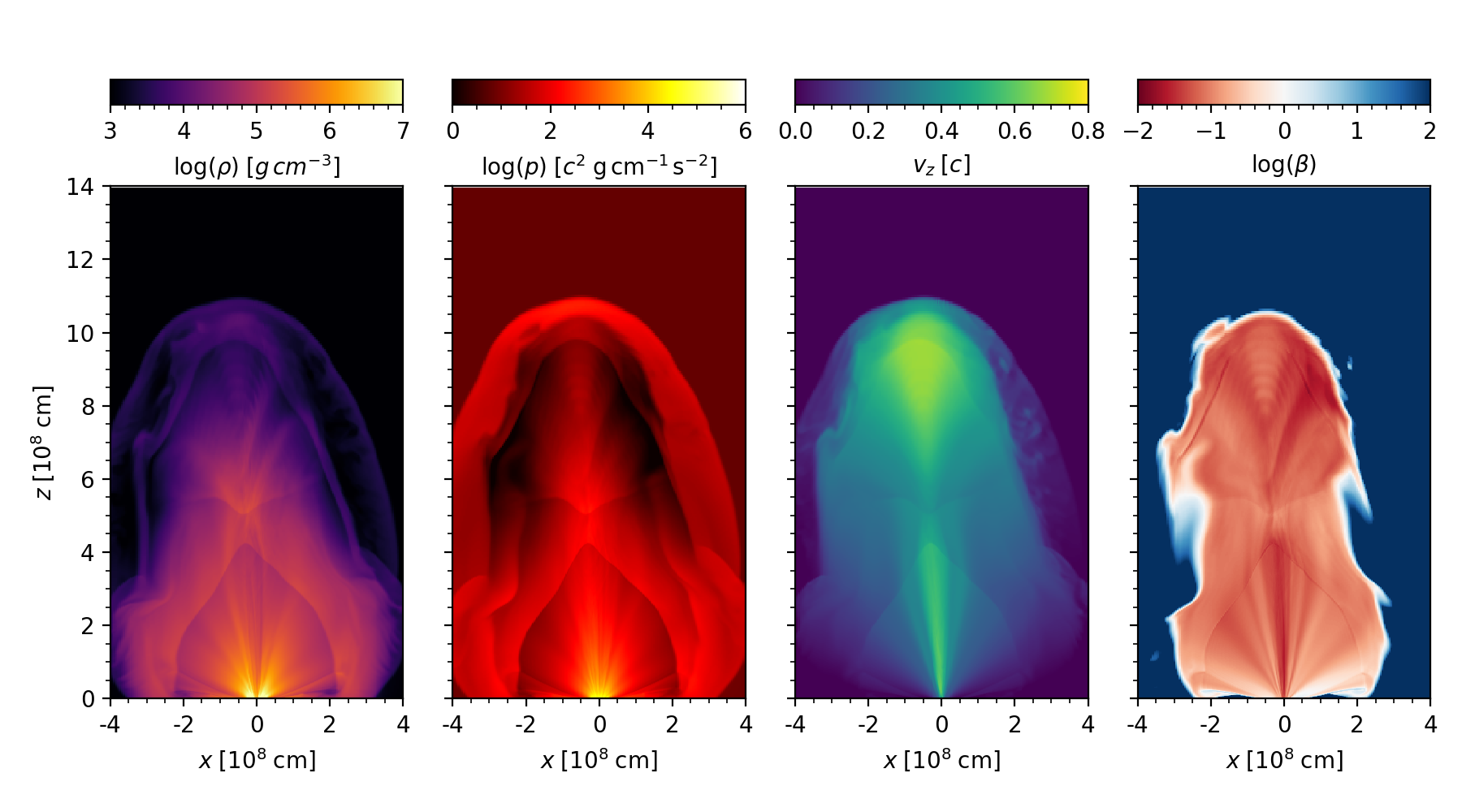}
    \caption{Meridional slice, $xz$-plane, of the density $\rho$, pressure $p$, velocity in the $z$-direction $v_z$ and plasma $\beta$, respectively from left to right, in the realistic jet at a late time in the simulation.}
    \label{fig:rd_multi}
\end{figure*}

The four panels in Figure \ref{fig:rd_multi} display, left to right, the gas density, pressure, velocity in the $z$-direction, and the ratio between gas and magnetic pressure $\beta = p_g/p_m$. This was taken at a late stage in our simulation.

The first two panels, showing respectively the gas density and pressure, do not present clear, collimated outflows, although this is clearly seen on the velocity panel, where the central part of the outflows is moving with higher velocities ($\sim 0.7\mathrm{c}$) compared to its surroundings. Furthermore, the plasma $\beta$ suggests a complex magnetic structure, which we discuss in Section \ref{sec:magfields}.

One noticeable aspect of the jet in this simulation, which can be clearly seen in Figure \ref{fig:rd_multi}, is that it gets pinched roughly between $z \sim 4\times 10^8\;\mathrm{cm}$ and $z \sim 5.5\times 10^8\;\mathrm{cm}$, where the presence of two shocks is clearly visible. These shocks appear much earlier on, as the outflows first start interacting with the surrounding material, and propagate as the simulation evolves. There, pressure and density both increase upon the bottom shock, while velocity decreases. In the top shock, around $z \sim 5.5\times 10^8\;\mathrm{cm}$, this behavior is reversed: density and pressure start to decrease, while the mean velocity starts to increase again. We also notice, near the head of the jet, the presence of further shocks associated with the interaction between the jet, the surrounding material, and the external medium. This is characterized by a sudden increase in both pressure and density, while the velocity slowly decreases downstream.

It should be noted that the jet in this simulation tilts a bit towards the left side in the $x$-direction (and also a little towards the right in the $y$-direction). Although this behavior is fairly normal in jets \citep{Mignone2013}, we cannot elaborate much on its future behavior, i.e., we cannot say whether this tilt would be reversed, or become even more pronounced. This is due to the very nature of our injection process, which consisted of $38$ snapshots being mapped into our grid in a constant loop. Hence, if the tilt was already forming as the snapshots were being injected, it is unlikely that its behavior would have changed in the spatial domain covered by this simulation. Evolving the jet for longer periods and distances could possibly lead to a further change in the jet behavior. Similarly, the injection of more data could also lead to a different behavior. Still, kink instabilities could lead to magnetic reconnection sites within the jet and subsequent particle acceleration along current sheets.

\subsection{Analytical jet}
\label{sec:results_aj}

\begin{figure*}
    \centering
    \includegraphics[width=\linewidth]{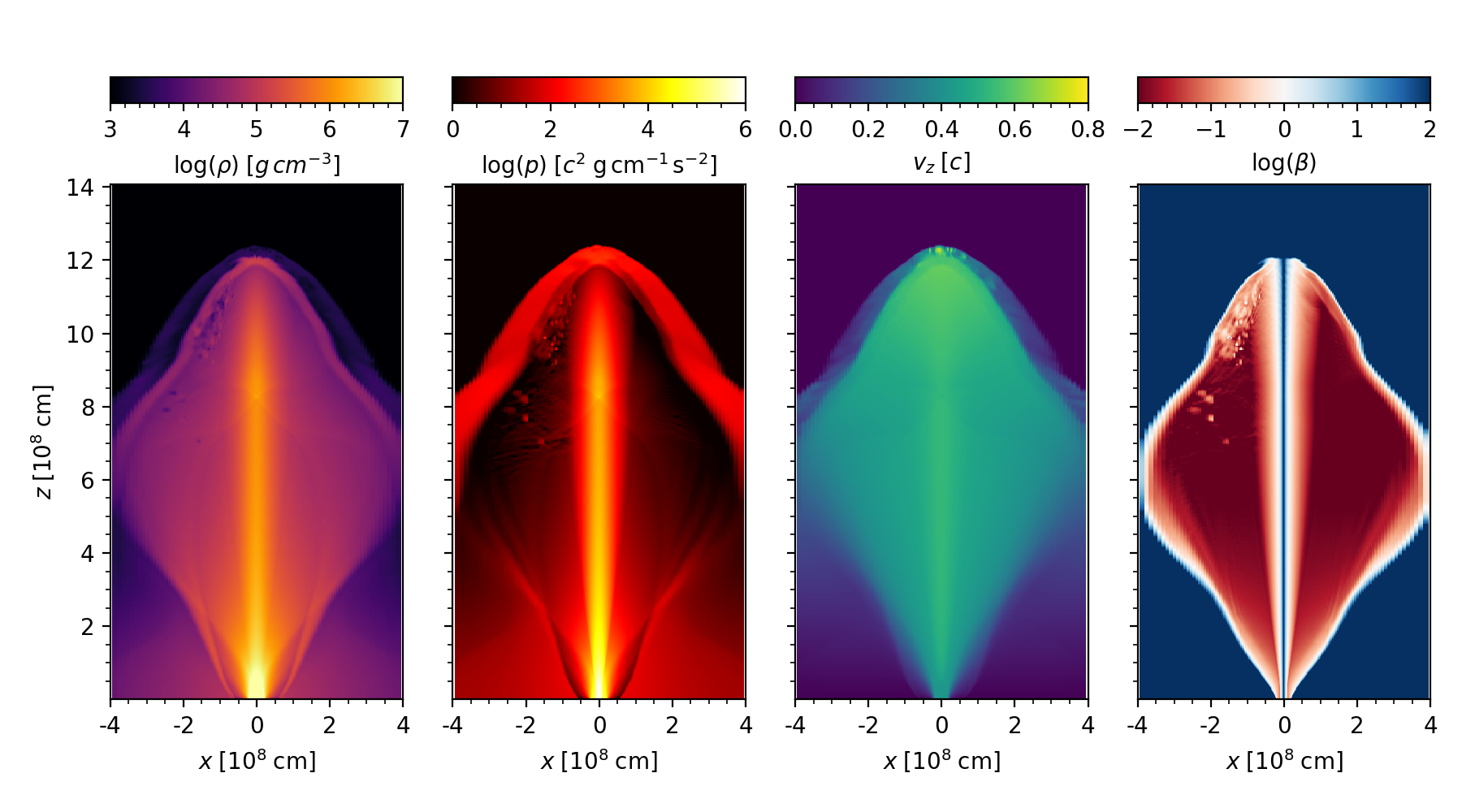}
    \caption{Meridional slice, $xz$-plane, of the density $\rho$, pressure $p$, velocity in the $z$-direction $v_z$ and plasma $\beta$, respectively from left to right, at a late time in the simulation, for the analytical jet initial data.}
    \label{fig:th_multi}
\end{figure*}

The analytical jet shows a qualitatively similar behavior on many aspects, when compared to the realistic jet. As seen on Figure \ref{fig:th_multi}, which is analogous to Figure \ref{fig:rd_multi}, the presence of a shock is visible around $8\times 10^8\;\mathrm{cm}$, although this is a very symmetric shock, with an X-shaped propagation, and the jet retains its overall form for the entire duration of the simulation, i.e., we found no significant pinching occurring along the jet. Having said this, we do notice a decrease in density and pressure along with an increase in velocity at $z \sim 8.2\times 10^8\;\mathrm{cm}$, which is similar to the behavior of the realistic jet.

As suggested by the plasma-$\beta$, the analytical jet appears to be far more collimated than the realistic jet. This could be due to our choice of the toroidal magnetization parameter $\sigma_\phi$, which was initialized at $\sigma_\phi = 1$. Overall, the jet retains a nearly axis-symmetrical structure throughout the entire simulation, and we found no signs pointing towards a change in that structure.

Furthermore, compared to the realistic jet, we see that the central axis of the analytical jet has a very clear high-$\beta$ region. Although the magnetic fields are comparable in both simulations, there is a striking difference in the pressure, as can be seen on the second panel of Figures \ref{fig:rd_multi} and \ref{fig:th_multi}, which accounts for this difference.

\section{Discussion}
\label{sec:discussion}

\subsection{Jet velocities and short GRBs}
\label{sec:velocities}

\begin{figure}
    \centering
    \includegraphics[width=\linewidth]{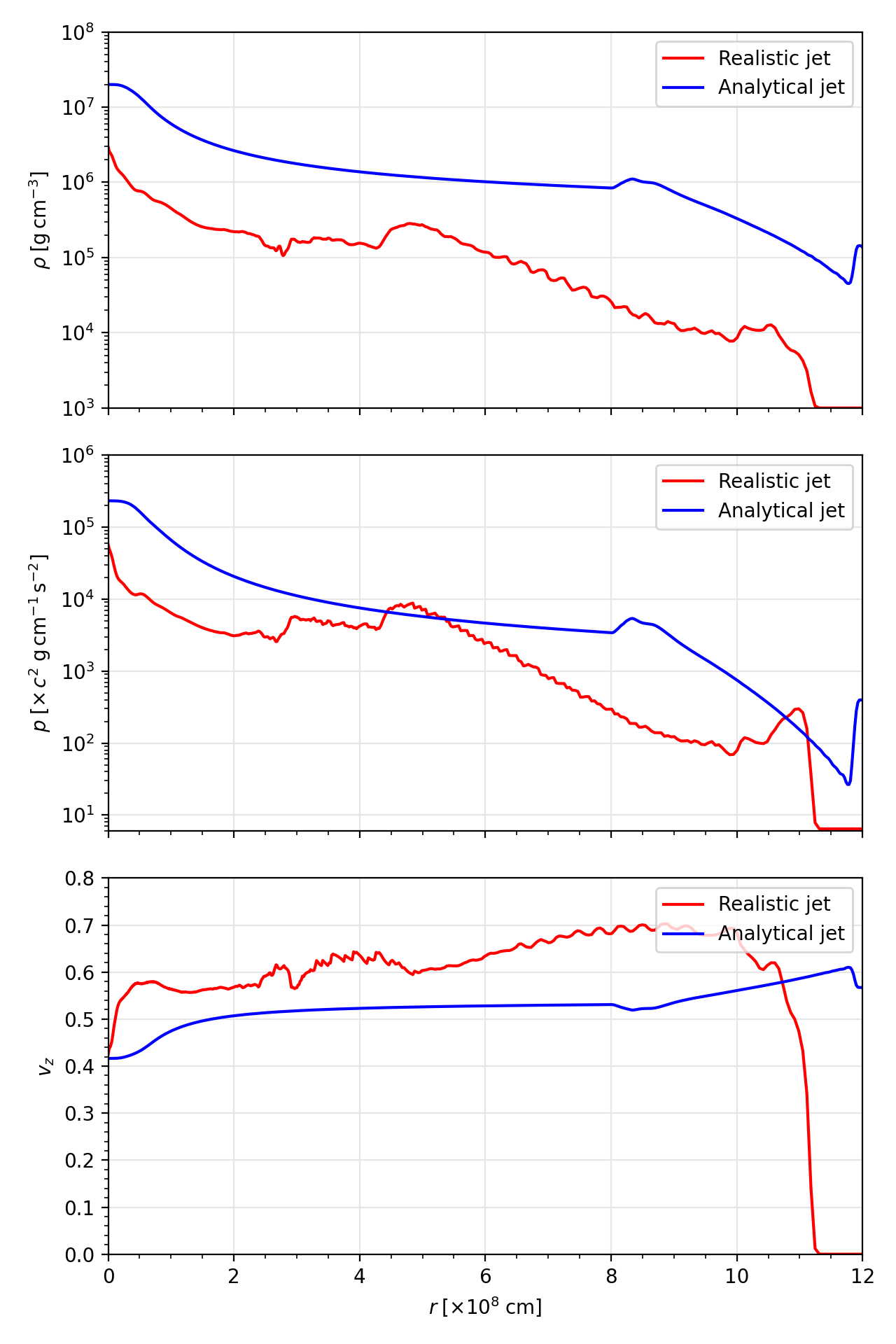}
    \caption{Profiles of the mean values of the density (top), pressure (center), and velocity (bottom) within $3.5\times 10^7\;\mathrm{cm}$ of the jet axis for the realistic jet (red) and analytical jet (blue) at a late stage.}
    \label{fig:profiles}
\end{figure}

The similarities in density and pressure between the two simulations are not unexpected, considering that our analytical jet had its initial values for these quantities chosen so that they would be similar to those in the realistic jet simulation. In order to better track and visualize these features, we traced the profiles of a few quantities, namely the density, pressure and velocity, along the jet. We did this by locating the jet axis -- taking into account its tilt -- and then taking the mean values of these quantities within a certain distance of the jet's central axis. This distance was obtained as follows: we first determined the jet axis by looking at points, on both $xz$- and $yz$-planes, where $\beta$ was lowest and $v_z$ was highest. We then determined, for each point in the axis, a circle along a plane perpendicular to it, to account for the jet being tilted. The circle radius is $\sim10^7\,\mathrm{cm}$, and was chosen by looking at the profiles for those quantities along cuts through these planes. This process was simplified for the analytical jet, as we did not have to deal with a tilted jet. The profiles are shown in Figure \ref{fig:profiles}. Top to bottom, they are the mean density, pressure and velocity along the portion of the outflows described above.

The features that we have qualitatively discussed before are better seen here. The shocks in the realistic jet do not appear to be very well-defined in Figure \ref{fig:profiles}, but this is simply because the shocks in this jet are ``pointy'' (see Figure \ref{fig:rd_multi}), marking the boundary between shocked and unshocked jet, and taking the average values of quantities within a small distance from the jet axis makes the shocks appear less obvious in Figure \ref{fig:profiles}.

The appearance of GRBs is associated with multiple factors. Crucial, though, are high speeds in the jet, while moderate to high speeds in the cocoon could also lead to sGRBs \citep{Gottlieb2018}. Since our engine powering the realistic jet is a neutron star before its collapse into a black hole, there are a few additional factors that affect the jet's properties that were taken into account in M20. For instance, neutrino cooling reduces baryon pollution in the polar regions, which in turn allows for the launch of faster jets. However, the main question still remains whether the jets are fast enough.

The velocities shown in Figure \ref{fig:profiles} are mean values within a small radius from the jet's central axis. We see that the jet reaches mean velocities of up to $\sim 0.71c$ in the realistic jet simulation, and $\lesssim 0.6c$ in the analytical jet. These correspond to Lorentz factors of $\Gamma \lesssim 1.45$, i.e., both jets only achieve moderate velocities, which suggests they are unsuitable as sGRB sources.

We note that, in M20, Lorentz factors of $\sim 2$ were found in the jet. Still, it is possible that those speeds were not maintained up to the moment of injection into our simulation, and the difference between those Lorentz factors and the ones we have here are not significant. Moreover, we can also estimate the Lorentz factor that could be achieved by our jet by using $\Gamma_\mathrm{max} = \Gamma_\mathrm{inj}(1+4p/\rho + \sigma)$. Our values of $\rho$, $p$ (see for example Fig. \ref{fig:profiles}) or $\beta$ (Fig. \ref{fig:rj_xy0}) do not suggest that such an increase would be substantial. Even if all magnetic energy could be used to power the jet, we would still obtain $\Gamma_\mathrm{max} < 5$ for the values of $\Gamma$, density and pressure encountered along the jet. 

\begin{figure}
    \centering
    \includegraphics[width=\linewidth]{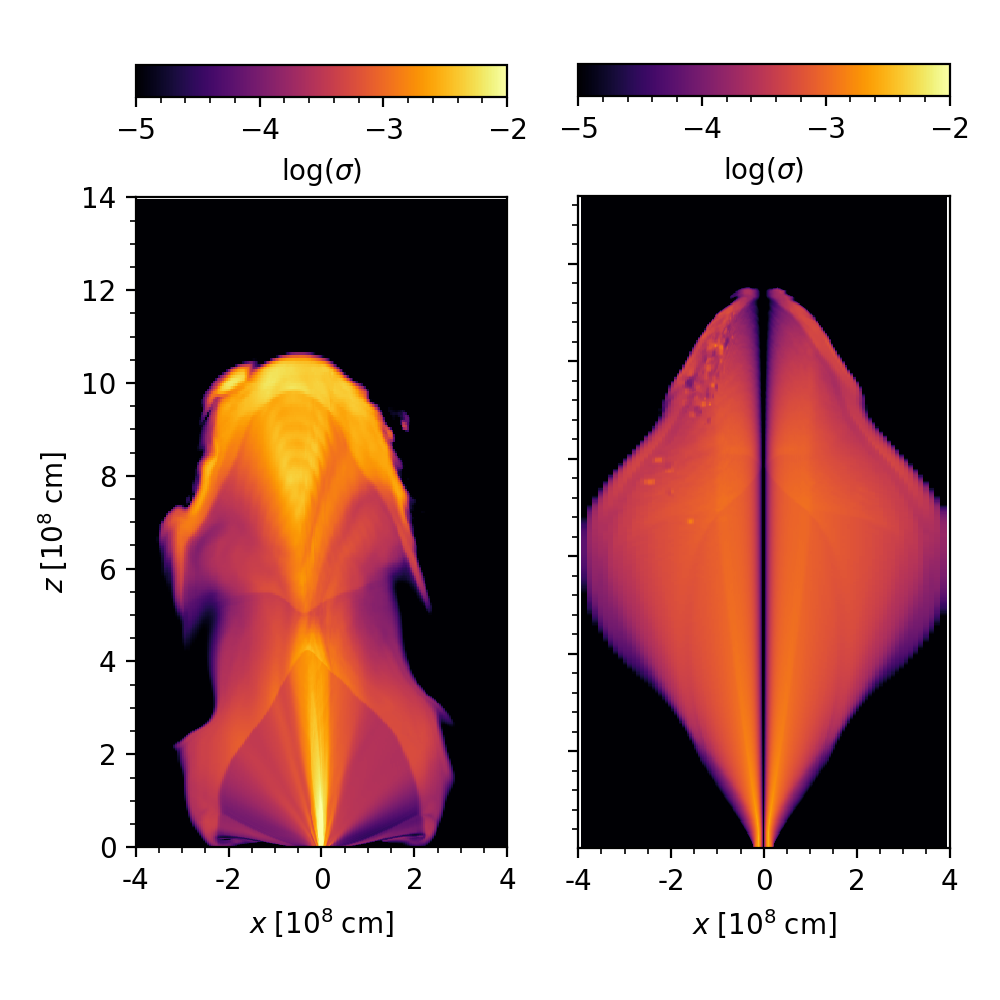}
    \caption{Meridional slice, $xz$-plane, of the magnetization $\sigma$ along the jet imported from M20 (left) and the analytical jet (right).}
    \label{fig:sigma}
\end{figure}

This can be seen in Figure \ref{fig:sigma}, which shows that the magnetization $\sigma$ does not reach high values in either jet. That said, we must factor in the effects of numerical constraints in our simulations. The simulation box in M20 was small, which in part motivated our decision to propagate their jet further, and even in this work we cannot extend it much beyond $10^9\,\mathrm{cm}$. This is due to both computational limitations and also the fact that we did not want to extend the looped injection of M20 jet into our grid, as the HMNS has already collapsed by the time our simulation ends and we cannot account for further changes in the jet that would occur once the BH is formed.

If such constraints are not present, we can argue that any limitations would now be related to the physics associated with the jets. Since we are propagating an already existing jet, it is reasonable to expect that any significant changes would have to be introduced at the early stages of a simulation, i.e., something analogous to M20. If a long-lived jet would be able to maintain its hollow nature, with comparatively low density and high pressure within its central axis, it could achieve higher Lorentz factors and therefore be a viable source of sGRBs. Furthermore, the analytical jet is limited by the way the initial quantities are prescribed, which is not the same as in the realistic jet. Even though both jets evolve differently, the initial velocity conditions are not conducive to reaching high Lorentz factors. Further propagation over at least another order of magnitude in distance where the background density drops significantly could, however, clarify the behavior of the jet and in particular its terminal speed.

With a hybrid approach, \cite{Pavan2021} evolved a top-hat jet within a realistic binary neutron star merger background, similar to \cite{Lazzati2021}, and showed that the jet achieves a terminal Lorentz factor of $\sim 40$. Furthermore, \cite{Gottlieb2022b} showed that a self-consistent jet launched upon a collapsar reached Lorentz factors of $\lesssim 30$ at distances of $10^{12}\;\mathrm{cm}$; comparatively, they propagate the jet 3 orders of magnitude more than in our evolution. \cite{Lazzati2021} also carry out a comparison with a fully analytical setup, like we have done in this work. Their findings are analogous to ours. The realistic and analytical simulations show comparable evolution but differ in some important details. In both cases, the realistic setup shows a less smooth jet, characterized by the presence of small structures in its velocity, density, pressure and --- in our case --- magnetization.

\cite{Gottlieb2018} presented a model in which a cocoon shock breakout could power GRB 170817A, provided the material in the cocoon achieved moderate to high velocities, $v \gtrsim 0.6-0.8c$. The total ejected mass in M20 is around $1.1\times 10^{-3}\;M_\odot$, and with the exception of the jet, none of the ejecta achieved velocities above $0.5c$. Upon propagating that jet, we found that the cocoon material still falls short of achieving velocities above $0.5c$. We found the same for the analytical jet. This further suggests that neither jet studied here is likely to lead to sGRBs in this scenario. However, the scenario proposed by \cite{Gottlieb2018} requires a tail of high velocity ejecta, with characteristic speed $\sim0.6-0.8c$. While such component could be dynamical in origin, it is also possible that it is provided during the metastable NS phase of the remnant. Our simulations show that, indeed, a collimated outflow with the required characteristics is 
produced by a magnetar jet.

The approximate neutrino leakage scheme, employed in M20, captures the overall neutrino energetics up to a factor of a few when compared to a full transport scheme in core-collapse supernovae simulations. The dependence on the energy, the deposition of momentum and neutrino pair-annihilation are not included and are possibly important in the jet formation process and the outflow properties. The approximate neutrino leakage scheme used in M20, thus, introduces some uncertainties and can be a potential source of errors.

It should be noted that the analytical simulation makes no direct assumptions regarding factors such as the neutron star EOS, neutrino effects or baryon pollution, all of which contribute to the NS lifetime, jet launching capabilities and jet velocity. Yet, the fact that both jets attained similar, moderate velocities, suggests that simply assuming a low/moderate initial velocity for the jet in the analytical simulation -- which in a realistic case would be due to the aforementioned factors -- partially makes up for the lack of explicit neutrino effects in this simulation. Nonetheless, the lack of relativistic speeds along the jets, or moderate-to-high speeds in the cocoon, makes them unlikely to produce sGRBs.

Even though the jet we study in this work does not move fast enough to power a short GRB jet, we notice that it remains collimated and does reach its terminal velocity, accelerating to its full potential. This makes it very interesting to explore, in a future follow-up work, of a magnetar jet with higher acceleration potential, for example due to higher magnetization or lower baryon loading.

\subsection{Magnetic field structure}
\label{sec:magfields}

The plots of $\beta$ in Figures \ref{fig:rd_multi} and \ref{fig:th_multi} suggest a very noticeable difference in the magnetic field structure of the two jets that we evolved. At first, this is somewhat expected, given the circumstances in which both simulations were started: the realistic jet is an evolution of data that had already been already subjected to previous magnetic field evolution, while the analytical jet was injected with a magnetic field profile given by Eq. \ref{eq:bmag}.

\begin{figure}
    \centering
    \includegraphics[width=\linewidth]{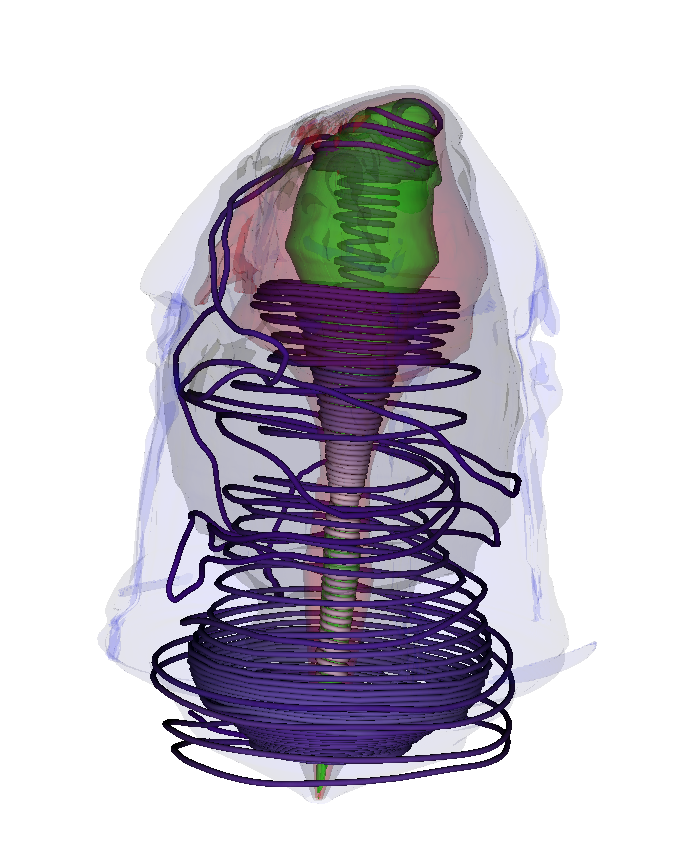}
    \caption{Contour plot of the $z$-component of the velocity, along with magnetic field lines in the realistic jet, showing the presence of both toroidal and poloidal components. Also shown are contour plots of the jet velocity in the $z$-direction, here taken at $v_z/c = 0.1, 0.25, 0.4, 0.55$.}
    \label{fig:mag_realistic}
\end{figure}

We show in Figure \ref{fig:mag_realistic} a contour plot of the $z$-component of the jet velocity along with the magnetic field lines. In M20 the purely-poloidal magnetic field initial configuration develops a strong toroidal component during the evolution. The field lines, in this case, retain their helical configuration even after propagating them into large distances.

\begin{figure}
    \centering
    \includegraphics[width=\linewidth]{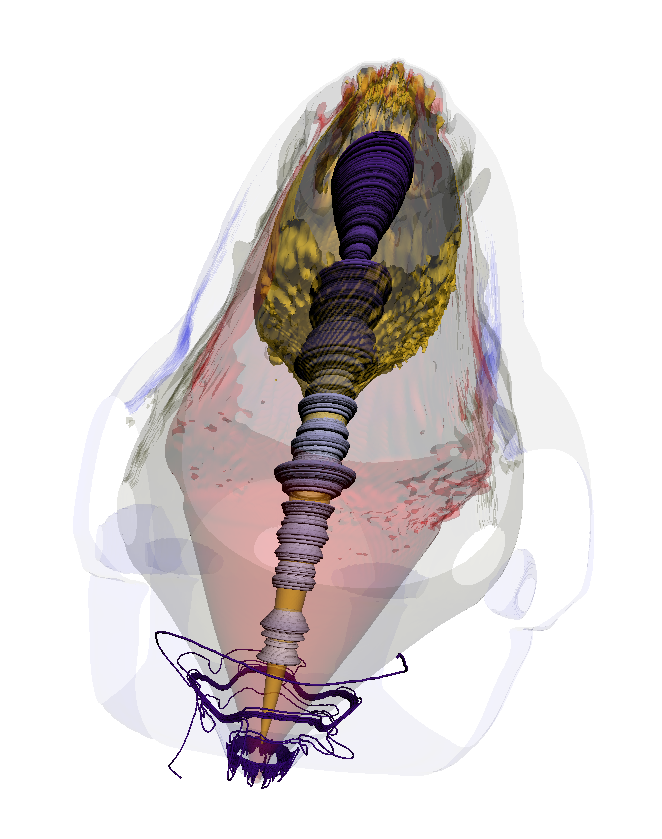}
    \caption{Same as Figure \ref{fig:mag_realistic}, but for the analytical jet. The ratio between poloidal and toroidal components in this jet is small, which required us to increase the number of integration steps to obtain the field lines shown here.}
    \label{fig:mag_analytical}
\end{figure}

The magnetic field of the analytical jet is shown in Figure \ref{fig:mag_analytical}, which is analogous to Figure \ref{fig:mag_realistic}. It was initialized here with a purely toroidal component (see Eq. \ref{eq:bmag}), and acquires a poloidal component as the jet propagates. Still, the ratio between its poloidal and toroidal components is smaller in this jet than in the realistic jet by at least a factor of $2$, leading to field lines with a compressed helical shape. From a numerical point of view, this only means that we had to increase significantly the amount of integration steps in order to obtain the field lines shown in Figure \ref{fig:mag_analytical}.

\subsection{Comparison with recent literature}
\label{sec:bhengine}

A fundamental difference between this work and most of the existing literature is that the jets we propagate here were launched by the neutron star before its collapse into a black hole. In other works (e.g. \citealt{Nathanail2020,Nathanail2021,Gottlieb2022a}), the jet component is driven by the black hole.

Additionally, general jet properties were discussed in \cite{MurguiaBerthier2017}. They found that a successful jet depends on a jet's head velocity being higher than that of the wind, which occurs in both M20 and here. The authors also argue that, since baryon pollution significantly decreases the maximum attainable Lorentz factor, a jet would only successfully emerge after the collapse into a BH.

Our NS-launched jet achieves slightly inferior Lorentz factors than those powered by black holes, but the fact that we have a successful jet raises the possibility of the material ejected during the HMNS phase playing a role in sweeping up the immediately surrounding material. Thus, it could pave the way for a strong jet component launched after the collapse into a black hole, especially if this is paired with a longer-lived jet-launching HMNS, whose longer steady-state operation could lead to higher terminal Lorentz factors. Yet, in the context of simulations, such quantities are somewhat arbitrary as they are fundamentally related to density floors and other ad hoc mechanisms added in order to guarantee a stable evolution.

Our assumption (Eq. \ref{eq:ejecta}) of a static ejecta is based on the semi-analytic model of \cite{Lazzati2019}. To test its validity and implications on the overall structure and motion of the jet (see \citealt{Hamidani2020,Hamidani2023,Gottlieb2022c} for the effects of a moving ejecta on the jet), we varied the ejecta velocity, in this model, between $0.01c$ and $0.3c$ in the NS frame. We found that the jet head velocity changes by $8\%$ ($0.65c$ in the slow case versus $0.7c$ in the fast case). Additionally, we found a variation of $6\%$ in the cocoon energy, the fast ejecta being more energetic. We consider these adequate if we take into account the uncertainties associated with the model, and as such, we consider our assumption of a static ejecta to be adequate in this context.

\section{Conclusions}
\label{sec:conclusions}

The main of aim of this work is to show the first steps towards a fully consistent end-to-end description of the BNS merger remnant as well as the jet launching and propagation. We use a realistic self-consistent jet, extracted from the GRMHD evolution of a highly-magnetized BNS merger remnant, to construct a mock long-lived steady-state jet. We employ this steady-state jet as initial data to address the problem of sGRB propagation through a BNS merger static medium. Additionally, as a baseline, we evolve an analytically prescribed jet in the same BNS merger medium. We discuss, for both cases, their ability to power short GRBs.

We find that both jets remain collimated and moderately magnetized and achieve moderate velocities, $v \sim 0.6-0.7c$ (Lorentz factors of$\sim 1.15-1.4$), and are surrounded by cocoons with speed $\lesssim 0.5c$. These velocities are close to the maximum attainable velocities of the injected outflows, demonstrating that the acceleration in the ejecta is efficient and that  the jet-ejecta interaction is not disruptive. We also find that the numerical, self consistent jet is slightly more efficient at reaching high velocity, by approximately $\sim20\%$. This small but significant difference between idealized and realistic simulations is consistent with what found by \cite{Lazzati2021}, who find a $\sim20\%$ effect when comparing short GRB jet evolution in realistic and idealized ambient materials. 

Due to the properties of the injected outflows, we cannot provide a definitive answer on whether jets from magnetars can or cannot power short GRBs, and the main reasons for this conclusion can be associated with numerical, rather than physical factors. The jets we study here have velocities large enough to contribute to the blue kilonova component seen in GW170817 and could help pre-evacuate a channel for the propagation of a subsequent relativistic jet. They do not possess, however, velocities large enough to attain a large Lorentz factor. The M20 simulations do show formation of fast outflows, with their high-resolution case (B15-high) leading to the highest Lorentz factors, $\Gamma \sim 5$. Nonetheless, the simulation we used here, B15-low, is sufficient to describe the large-scale dynamics and propagation of the jet.

Because the flow's maximum speed seems to be limited by numerical constraints (e.g. resolution), there is hope that future general-relativistic calculations would provide input of potentially faster jets, either by increased magnetic field or entropy, or thanks to lower baryon contamination. In addition, as seen in the simulation B15-high in M20, resolving the MRI-driven turbulence plays a fundamental role in jet emission, achievable Lorentz factors, and lifespan of the remnant, as the HMNS in B15-high lives 7ms longer than the low-resolution counterpart, an increase of $\sim20\%$. Furthermore, neutrino effects also play an important role in jet emission in the GRMHD evolution of the remnant, and so the inclusion of neutrino pair-annihilation may boost the jet's Lorentz factor to the relativistic sGRB regime. If these matters are addressed, it appears reasonable that a simulation where a GRB is powered by a magnetar can be achieved.

\acknowledgements We thank the anonymous referee for their comments, which improved the manuscript. GS and DL acknowledge support from NSF grant AST-1907955. Some simulations were carried out on NCSA's BlueWaters allocation ILL\_baws, TACC's Frontera under allocation DD FTA-Moesta, and LRZ's SuperMUC-NG under PRACE award 2021250017 allocation pn49ju. Resources supporting this work were provided by the NASA High-End Computing (HEC) Program through the NASA Advanced Supercomputing (NAS) Division at Ames Research Center. Computational resources were also provided by the CoSINe cluster at Oregon State University.

\bibliographystyle{aasjournal}
\bibliography{refs2}

\end{document}